\documentstyle[12pt,epsf]{article}

\let\huge=\Large
\let\Large=\large
\let\large=\normalsize
\newcommand{\Preprint}{\vspace*{-2.6cm} \noindent hep-ph/9509253 \hfill
  FTUV/95-43 \\ \mbox{}\hfill IFIC/95-45 \\  \mbox{}\hfill
  September 1995\vspace{0.3cm}}
\def\Thebibliography#1{\section*{\centering REFERENCES}\list
{\arabic{enumi}.}{\settowidth\labelwidth{#1.}\leftmargin\labelwidth
 \advance\leftmargin\labelsep
 \usecounter{enumi}}
 \def\newblock{\hskip .11em plus .33em minus .07em}
 \sloppy\clubpenalty4000\widowpenalty4000
 \sfcode`\.=1000\relax}

\def\refjl#1#2#3#4#5#6{\bibitem{#1} #2, {\it #3 \/} {\bf #4} (#5) #6.}
\def\etal{{\it et al\/}}
\def\NP{Nucl. Phys.}
\def\PL{Phys. Lett.}
\def\PRL{Phys. Rev. Lett.}
\def\PR{Phys. Rev.}
\def\PRep{Phys. Rep.}
\def\ZP{Z. Phys.}

\def\APNY{Ann. Phys. (NY)}

\newcommand{\eqn}[1]{(\ref{#1})}
\newcommand{\bel}[1]{\be\label{#1}}
\newcommand{\be}{\begin{equation}}
\newcommand{\ee}{\end{equation}}
\newcommand{\ba}{\begin{array}{c}}
\newcommand{\ea}{\end{array}}
\newcommand{\beqn}{\begin{eqnarray}}
\newcommand{\eeqn}{\end{eqnarray}}

\newcommand{\bi}{\begin{itemize}}
\newcommand{\ei}{\end{itemize}}

\newcommand{\cP}{{\cal P}}

\newcommand{\no}{\nonumber}

\newcommand{\rms}{\rm\scriptsize}
\newcommand{\tcf}{$\tau$cF}
\begin{document}

\newpage  
\Preprint
\begin{center}
\renewcommand{\thefootnote}{\fnsymbol{footnote}}
   {\huge\bf IMPORTANCE OF PRECISION MEASUREMENTS IN THE TAU
    SECTOR\footnote{Talk given at the \tcf\ workshop, Argonne (Illinois),
    USA, June 21-23, 1995}}\addtocounter{footnote}{-1}
   \\[2ex]
  {{\Large\bf A. Pich} \\[1.5ex] {\it
  Departament de F\'{\i}sica Te\`orica, IFIC,  
  Universitat de Val\`encia -- CSIC \\ E-46100
  Burjassot, Val\`encia, Spain}}
   \\[2ex]
\parbox[t]{12.5cm}{\small
{\bf Abstract.}
$\tau$ decays provide a powerful tool to test the structure
of the weak currents and the universality of their couplings to
the $W$ boson.
The constraints implied by present data
and the possible
improvements at the \tcf\ are analyzed.}
\end{center}

%
%
\section*{\centering INTRODUCTION}
\label{sec:introduction}

The light quarks and leptons are by far the best known ones.
Many experiments have
analyzed in the past the properties of
$e$, $\mu$, $\nu_e$, $\nu_\mu$, $\pi$, $K$, $\ldots$
However, one na\"{\i}vely
expects the heavier fermions to be much more sensitive to New Physics,
since they may couple more strongly to whatever dynamics is
responsible for the fermion-mass generation.
Obviously, new heavy-flavour facilities,
such as the $B$ and Tau-Charm Factories (\tcf ), are needed
to  match (at least) the precision attained for the light flavours.

Similarly to the bottom quark, the tau lepton is a third
generation  fermion, with a wide variety of decay channels into
particles belonging to the first and second fermionic families.
Therefore, one can expect that
$\tau$ and $b$ physics will provide some clues to the puzzle of
the recurring generations of leptons and quarks.
While the decays of the $b$-quark are ideally suited to look for
quark mixing and CP-violating phenomena,
the pure leptonic or semileptonic character
of $\tau$ decays provides a much  cleaner laboratory to test the
structure
of the weak currents and the universality of their couplings to
the gauge bosons.
Moreover, the tau is the only known lepton massive enough to decay
into hadrons; its semileptonic decays are then an ideal
tool for studying strong interaction effects in very clean conditions.

The last five years have witnessed a substantial change on our
knowledge of the $\tau$ properties.
The large (and clean) data samples collected by the most recent
experiments have
improved considerably the statistical accuracy and, moreover,
have brought
a new level of systematic understanding.
The qualitative change of the $\tau$ data can be appreciated in
Table~\ref{tab:improvements}, which compares the status of several
$\tau$
measurements in 1990 \cite{PDG:90,PI:92} with the most recent
world averages \cite{PDG:94,montreux}.
All experimental results obtained so far
confirm the Standard Model (SM) scenario
in which the $\tau$ is a sequential
lepton, with its own quantum number and associated neutrino.

\begin{table}[bth]
\caption{Recent improvements in $\tau$ physics and expected
precision at the \tcf .}
\label{tab:improvements}
\centering\vspace{0.2cm}
\begin{tabular}{|c|c|c|c|}
\hline
Parameter & 1990 \protect\cite{PDG:90,PI:92} &
  1995 \protect\cite{PDG:94,montreux} & \tcf\ sensitivity
\\ \hline
$m_\tau$ \, (MeV) & $1784.1^{+2.7}_{-3.6}$ & $1777.0\pm 0.3$ & $0.1$
\\
$m_{\nu_\tau}$ \, (MeV) & $< 35$ \quad {\footnotesize (a)} &
$< 24$ \quad {\footnotesize (a)} & 1--2
\\
$\tau_\tau$ \, (fs) & $303\pm 8$ & $291.6\pm 1.6$ & --
\\
$B_e$ \, (\%) & $17.7\pm 0.4$ & $17.79\pm 0.09$ & 0.02
\\
$B_\mu$ \, (\%) & $17.8\pm 0.4$ & $17.33\pm 0.09$ & 0.02
\\
$B(\pi^-\nu_\tau)$ \, (\%) & $11.0\pm 0.5$ & $11.09\pm 0.15$ & 0.01
\\
$B(K^-\nu_\tau)$ \, (\%) & $0.68\pm 0.19$ & $0.68\pm 0.04$ & 0.003
\\
$B(\pi^-\eta\nu_\tau)$ & $<9\times 10^{-3}$  \quad {\footnotesize (a)}
& $<3.4\times 10^{-4}$
\quad {\footnotesize (a)} & $10^{-6}$
\\
$B(l^-G)$ & $<10^{-2}$ & $<2.7\times 10^{-3}$
\quad {\footnotesize (a)} & $10^{-5}$
\\
$B(\mu^-\gamma)$ & $< 5.5\times 10^{-4}$ \quad {\footnotesize (b)} &
  $< 4.2\times 10^{-6}$ \quad {\footnotesize (b)} & $10^{-7}$
\\
$B(e^-e^+e^-)$ & $< 3.8\times 10^{-5}$ \quad {\footnotesize (b)} &
  $< 3.3\times 10^{-6}$ \quad {\footnotesize (b)} & $10^{-7}$
\\ \hline
$\rho_{\tau\to\mu}$ & $0.84\pm 0.11$ & $0.738\pm 0.038$ & 0.002
\\
$\eta_{\tau\to\mu}$ & -- & $-0.14\pm 0.23$ & 0.003
\\
$\xi_{\tau\to\mu}$ & -- & $1.23\pm 0.24$ & 0.02
\\
$(\xi\delta)_{\tau\to\mu}$ & -- & $0.71\pm 0.15$ & 0.02
\\
$\xi'_{\tau\to\mu}$ & -- & -- & 0.15
\\
$h_{\nu_\tau}$ & -- & $-1.014\pm0.027$ & 0.003
\\ \hline
$a_\tau^\gamma$ & $<0.1$ \quad {\footnotesize (b)} &
$<0.01$ \quad {\footnotesize (a)} & 0.001
\\
$d_\tau^\gamma$ \, (e cm) & $<6\times 10^{-16}$ \quad {\footnotesize (b)}&
$<5\times 10^{-17}$ \quad {\footnotesize (a)} & $10^{-17}$
\\ \hline
\end{tabular}
\vspace{0.2 cm}
{\footnotesize (a) 95\% CL \ ; \quad  (b) 90\% CL}
\end{table}

The present experiments are soon going to reach their
systematic limits. Further improvements in $\tau$ physics require
then new high-precision facilities,
to push the significance of the $\tau$ tests beyond the present
few per cent level.
The last column in Table~\ref{tab:improvements} shows the
sensitivities that could be achieved at the \tcf .
In some cases, a much better accuracy could be obtained
with polarized beams or monochromatic optics.

In the following, I discuss several precision tests
of the SM, using the present $\tau$-decay data,
and the expected improvements at the \tcf .
I will concentrate on the universality and Lorentz-structure
of the charged leptonic currents.
A discussion of other important topics in $\tau$ physics can be
found in refs.~[2, 5--8].

\section*{\centering CHARGED-CURRENT UNIVERSALITY}

The leptonic decays
$\tau^-\to e^-\bar\nu_e\nu_\tau,\mu^-\bar\nu_\mu\nu_\tau$
are theoretically understood at the level of the electroweak
radiative corrections \cite{MS:88}.
Within the SM,   
\begin{equation}
\label{eq:leptonic}
\Gamma_{\tau\to l} \, \equiv \,
\Gamma (\tau^- \rightarrow \nu_{\tau} l^- \bar{\nu}_l)  \, = \,
  {G_F^2 m_{\tau}^5 \over 192 \pi^3} \, f(m_l^2 / m_{\tau}^2) \,
r_{EW},
\end{equation}
where $f(x) = 1 - 8 x + 8 x^3 - x^4 - 12 x^2 \log{x}$.
The factor $r_{EW}=0.9960$ takes into account radiative corrections
not included in the
Fermi coupling constant $G_F$, and the non-local structure of the
$W$ propagator \cite{MS:88}.

Using the value of $G_F$
measured in $\mu$ decay, Eq.~\eqn{eq:leptonic}
provides a
relation \cite{PI:92} between the $\tau$ lifetime
and the leptonic branching ratios
$B_l\equiv B(\tau^-\to\nu_\tau l^-\bar\nu_l)$:
\begin{equation}
\label{eq:relation}
B_e \, = \, {B_\mu \over 0.972564\pm 0.000010} \, =
{ \tau_{\tau} \over (1.6321 \pm 0.0014) \times 10^{-12}\, {\rm s} } \, .
\end{equation}
The errors reflect the present uncertainty of $0.3$ MeV
in the value of $m_\tau$.

\begin{figure}[bht]
\centerline{\epsfxsize =5.7in \epsfbox{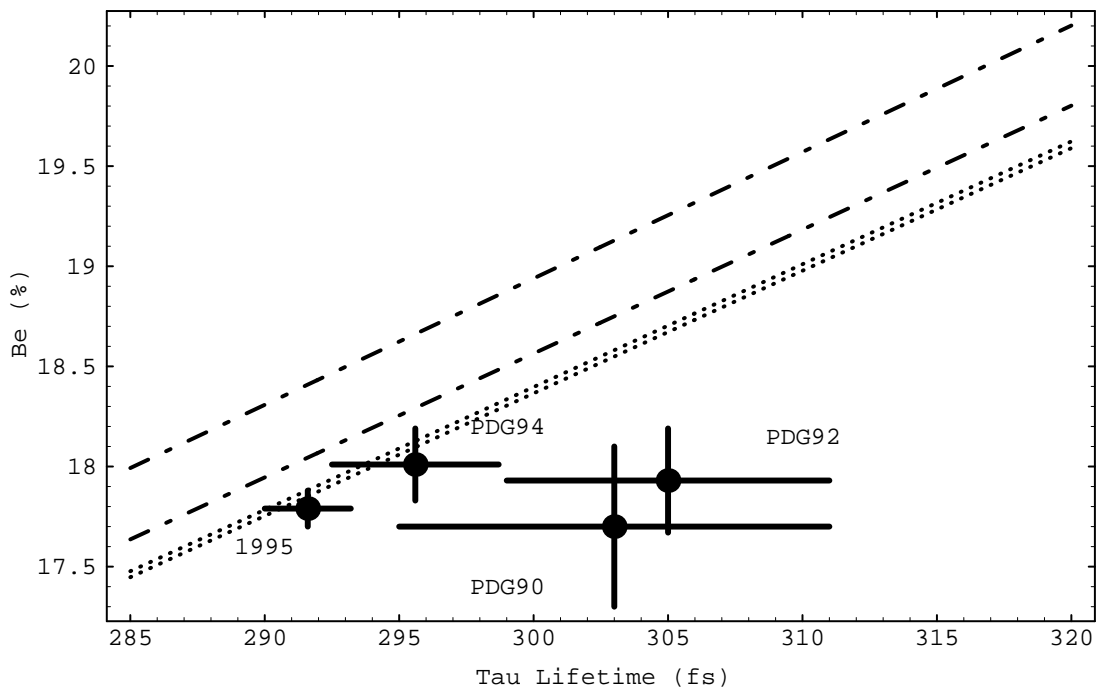}}
\caption{Relation between $B_e$ and $\tau_\tau$. The narrow dotted
band corresponds to the prediction in Eq.~(\protect\ref{eq:relation}).
The larger region between the two dot-dashed lines indicates
the relation obtained with the old \protect\cite{PDG:92}
value of $m_\tau$.
The experimental points show the present world averages
\protect\cite{montreux},
together with the values quoted by the Particle Data Group
[1, 3, 10]
since 1990.}
\label{fig:BeLife}
\end{figure}

The predicted $B_\mu/B_e$ ratio is in perfect agreement with the measured
value $B_\mu/B_e = 0.974 \pm 0.007$ \cite{montreux}.  As shown in
Fig.~\ref{fig:BeLife}, the relation between $B_e$ and
$\tau_\tau$ is also well satisfied by the present data. Notice,  that this
relation is very sensitive to the value of the $\tau$ mass
[$\Gamma_{\tau\to l}\propto m_\tau^5$]. The most recent measurements of
$\tau_\tau$, $B_e$ and $m_\tau$ have consistently moved the world averages
in the correct direction, eliminating the previous ($\sim 2\sigma$)
disagreement. The experimental precision (0.5\%) is already approaching the
level where a possible non-zero $\nu_\tau$ mass could become relevant; the
present bound \cite{ALEPH:95}
$m_{\nu_\tau}< 24$ MeV (95\% CL) only guarantees that such effect is below
0.14\%.

These measurements can be used to test the universality of
the $W$ couplings to the leptonic charged currents.
The $B_\mu/B_e$ ratio constraints $|g_\mu/g_e|$, while the
$B_e/\tau_\tau$ relation provides information on $|g_\tau/g_\mu|$.
The present results are shown in Tables \ref{tab:univme} and
\ref{tab:univtm}, together with the values obtained from the
ratio \cite{BR:92}   
$R_{\pi\to e/\mu}\equiv\Gamma(\pi^-\to e^-\bar\nu_e)/
\Gamma(\pi^-\to\mu^-\bar\nu_\mu)$,
and from the comparison of the $\sigma\cdot B$ partial production
cross-sections for the various $W^-\to l^-\bar\nu_l$ decay
modes at the $p$-$\bar p$ colliders \cite{UA1:89}. 

\begin{table}[bt]
\caption{Present constraints on $|g_\mu/g_e|$.}
\label{tab:univme}
\centering\vspace{0.2cm}
\begin{tabular}{|c|c|c|c|}
\hline
& $B_\mu/B_e$ \cite{montreux} & $R_{\pi\to e/\mu}$ \cite{BR:92}  %
&
$\sigma\cdot B_{W\to\mu/e}$ \cite{UA1:89} 
\\ \hline
$|g_\mu/g_e|$ & $1.0008\pm 0.0036$ & $1.0017\pm 0.0015$ &
$1.01\pm 0.04$
\\ \hline
\end{tabular}
\end{table}
\begin{table}[bt]
\caption{Present constraints on $|g_\tau/g_\mu|$.}
\label{tab:univtm}
\centering\vspace{0.2cm}
\begin{tabular}{|c|c|c|c|c|}
\hline
& $B_e\tau_\mu/\tau_\tau$ \cite{montreux} &
$R_{\tau/\pi}$ \cite{montreux} & $R_{\tau/K}$ \cite{montreux} &
$\sigma\cdot B_{W\to\tau/\mu}$ \cite{UA1:89} 
\\ \hline
$|g_\tau/g_\mu|$ & $0.9979\pm 0.0037$ & $1.006\pm 0.008$ &
$0.972\pm 0.029$ & $0.99\pm 0.05$
\\ \hline
\end{tabular}
\end{table}

The decay modes $\tau^-\to\nu_\tau\pi^-$ and $\tau^-\to\nu_\tau K^-$
can also be used to test universality through the ratios
\beqn\label{eq:R_tp}
R_{\tau/\pi} & \!\!\!\equiv &\!\!\!
 {\Gamma(\tau^-\to\nu_\tau\pi^-) \over
 \Gamma(\pi^-\to \mu^-\bar\nu_\mu)} =
\Big\vert {g_\tau\over g_\mu}\Big\vert^2
{m_\tau^3\over 2 m_\pi m_\mu^2}
{(1-m_\pi^2/ m_\tau^2)^2\over
 (1-m_\mu^2/ m_\pi^2)^2}
\left( 1 + \delta R_{\tau/\pi}\right) , \qquad
\\ \label{eq:R_tk}
R_{\tau/K} &\!\!\! \equiv &\!\!\! {\Gamma(\tau^-\to\nu_\tau K^-) \over
 \Gamma(K^-\to \mu^-\bar\nu_\mu)} =
\Big\vert {g_\tau\over g_\mu}\Big\vert^2
{m_\tau^3\over 2 m_K m_\mu^2}
{(1-m_K^2/m_\tau^2)^2\over
(1-m_\mu^2/ m_K^2)^2}
\left( 1 + \delta R_{\tau/K}\right) , \qquad
\eeqn
where the dependence on the hadronic matrix elements (the so-called
decay constants $f_{\pi,K}$) factors out.
Owing to the different energy scales involved, the radiative
corrections to the $\tau^-\to\nu_\tau\pi^-/K^-$ amplitudes
are however not the same than the corresponding effects in
$\pi^-/K^-\to\mu^-\bar\nu_\mu$. The size of the relative
correction was first estimated by
Marciano and Sirlin \cite{MS:93} to be
$\delta R_{\tau/\pi} = (0.67\pm 1.)\% $, where the 1\% error amounts
for the missing long-distance contributions to the
tau decay rate.
A recent evaluation of those long-distance corrections \cite{DF:94}
quotes the more precise values:
\bel{eq:dR_tp_tk}
\delta R_{\tau/\pi} = (0.16\pm 0.14)\% \ ,    \qquad\qquad
\delta R_{\tau/K} = (0.90\pm 0.22)\%  \ .
\ee
Using these numbers, the measured \cite{montreux}
$\tau^-\to\pi^-\nu_\tau$
and $\tau^-\to K^-\nu_\tau$ decay rates imply the
$|g_\tau/g_\mu|$ ratios given in Table~\ref{tab:univtm}.
The inclusive sum of both decay modes, i.e.
$\Gamma[\tau^-\to h^-\nu_\tau]$ with $h=\pi,K$, provides a
slightly more accurate determination:
$|g_\tau/g_\mu| = 1.004\pm 0.007$.

The present data verifies the universality of the leptonic
charged-current couplings to the 0.15\% ($e/\mu$) and 0.37\%
($\tau/\mu$) level. The precision of the most recent
$\tau$-decay measurements is becoming competitive with the
more accurate $\pi$-decay determination.
It is important to realize the complementarity of the
different universality tests.
The pure leptonic decay modes probe
the charged-current couplings of a transverse $W$. In contrast,
the decays $\pi/K\to l\bar\nu$ and $\tau\to\nu_\tau\pi/K$ are only
sensitive to the spin-0 piece of the charged current; thus,
they could unveil the presence of possible scalar-exchange
contributions with Yukawa-like couplings proportional to some
power of the charged-lepton mass.
One can easily imagine new-physics scenarios which would modify
differently the two types of leptonic couplings \cite{MA:94}.
For instance,
in the usual two-Higgs doublet model, charged-scalar exchange
generates a correction to the ratio $B_\mu/B_e$, but
$R_{\pi\to e/\mu}$ remains unaffected.
Similarly, lepton mixing between the $\nu_\tau$ and an hypothetical
heavy neutrino would not modify the ratios  $B_\mu/B_e$ and
$R_{\pi\to e/\mu}$, but would certainly correct the relation between
$B_l$ and the $\tau$ lifetime.

At the \tcf, the accurate measurement of the $B_\mu/B_e$ ratio
would allow to test $|g_\mu/g_e|$ to the 0.05\% level, compared
to the present 0.36\% precision (0.15\% from $R_{\pi\to e/\mu}$).
The final accuracy of the $|g_\tau/g_\mu|$ universality test
will be limited by the knowledge of the $\tau$ lifetime.
Assuming that the $\tau_\tau$ measurement will be improved
(at LEP or at the $B$ Factory) by
a factor of 2, i.e.
$\delta\tau_\tau/\tau_\tau\sim 0.3\%$,
$|g_\tau/g_\mu|$ would be tested with a 0.16\% precision.

\section*{\centering LORENTZ STRUCTURE OF THE CHARGED CURRENT}

Let us consider the leptonic decays $l^-\to\nu_l l'^-\bar\nu_{l'}$,
where the lepton pair ($l$, $l^\prime $)
may be ($\mu$, $e$), ($\tau$, $e$), or ($\tau$, $\mu$).
The most general, local, derivative-free, lepton-number conserving,
four-lepton interaction Hamiltonian,
consistent with locality and Lorentz invariance
[17--23],
%
\be
{\cal H} = 4 \frac{G_{l'l}}{\sqrt{2}}
\sum_{\epsilon,\omega = R,L}^{n = S,V,T}
g^n_{l'_\epsilon l^{\phantom{'}}_\omega}
\left[ \overline{l'_\epsilon}
\Gamma^n {(\nu_{l'})}_\sigma \right]\,
\left[ \overline{({\nu_l})_\lambda} \Gamma_n
	l_\omega \right]\ ,
\label{eq:hamiltonian}
\ee
contains ten complex coupling constants or, since a common phase is
arbitrary, nineteen independent real parameters.
$\epsilon , \omega , \sigma, \lambda$ are the chiralities (left-handed,
right-handed)  of the  corresponding  fermions, and $n$ labels the
type of interaction:
scalar ($I$), vector ($\gamma^\mu$), tensor
($\sigma^{\mu\nu}/\sqrt{2}$).
For given $n, \epsilon ,
\omega $, the neutrino chiralities $\sigma $ and $\lambda$
are uniquely determined.
Taking out a common factor $G_{l'l}$, which is determined by the total
decay rate, the coupling constants $g^n_{l'_\epsilon l_\omega}$
are normalized to \cite{FGJ:86}
\beqn\label{eq:normalization}
1 &\!\!\! = &\!\!\!
{1\over 4} \,\left( |g^S_{RR}|^2 + |g^S_{RL}|^2
    + |g^S_{LR}|^2 + |g^S_{LL}|^2 \right)
  \, + \, 3 \,\left( |g^T_{RL}|^2 + |g^T_{LR}|^2 \right)
\no \\ & &\!\!\! \mbox{}
+ \left(
   |g^V_{RR}|^2 + |g^V_{RL}|^2 + |g^V_{LR}|^2 + |g^V_{LL}|^2 \right)
\, .
\eeqn
In the SM, $g^V_{LL}  = 1$  and all other
$g^n_{\epsilon\omega} = 0 $.

For an initial lepton-polarization ${\cal P}_l$,
the final charged lepton distribution in the decaying lepton
rest frame
is usually parametrized in the form  \cite{BM:57,KS:57}
\begin{eqnarray}\label{eq:spectrum}
{d^2\Gamma \over dx\, d\cos\theta} &\!\!\! = &\!\!\!
{m_l\omega^4 \over 2\pi^3} G_{l'l}^2 \sqrt{x^2-x_0^2}\,
  \Biggl\{ x (1 - x) + {2\over 9} \rho
 \left(4 x^2 - 3 x - x_0^2 \right)
+  \eta\, x_0 (1-x)
\Biggr.\nonumber\\ & & \Biggl.
  - {1\over 3}{\cal P}_l \, \xi \, \sqrt{x^2-x_0^2} \cos{\theta}
  \left[ 1 - x + {2\over 3}  \delta \left( 4 x - 4 + \sqrt{1-x_0^2}
\right)\right]  \Biggr\} \, , \quad
\end{eqnarray}
where $\theta$ is the angle between the $l^-$ spin and the
final charged-lepton momentum,
$\, \omega \equiv (m_l^2 + m_{l'}^2)/2 m_l \, $
is the maximum $l'^-$ energy for massless neutrinos, $x \equiv E_{l'^-} /
\omega$ is the reduced energy and $x_0\equiv m_{l'}/\omega$.
For unpolarized $l's$, the distribution is characterized by
the so-called Michel \cite{MI:50} parameter $\rho$
and the low-energy parameter $\eta$. Two more parameters, $\xi$
and $\delta$ can be determined when the initial lepton polarization is known.
If the polarization of the final charged lepton is also measured,
5 additional independent parameters \cite{PDG:94}
($\xi'$, $\xi''$, $\eta''$, $\alpha'$, $\beta'$)
appear.

The total decay rate is given by
(neutrinos are assumed to be massless)
\be\label{eq:gamma}
\Gamma\, = \, {m_l^5 G_{l'l}^2\over 192 \pi^3}\,
\left\{ f\!\left({m_{l'}^2\over m_l^2}\right)
+ 4\eta\, {m_{l'}\over m_l}\, g\!\left({m_{l'}^2\over m_l^2}\right)
\right\} r_{\mbox{\rms EW}}
\, ,
\ee
where
$g(z) = 1 + 9 z - 9 z^2 - z^3 + 6 z (1+z) \ln{z}$,
and the SM
radiative corrections $r_{\mbox{\rms EW}}$
have been included\footnote{
Since we assume that the SM provides the dominant
contribution
to the decay rate, any additional higher-order correction
beyond the effective four-fermion Hamiltonian
(\protect\ref{eq:hamiltonian})
would be a subleading effect.}.

Thus, the normalization $G_{e\mu}$ corresponds to the Fermi coupling
$G_F$, measured in $\mu$ decay.
The $B_\mu/B_e$ and $B_e\tau_\mu/\tau_\tau$
universality tests, discussed in the previous section,
actually prove the ratios
$|\hat G_{\mu\tau}/\hat G_{e\tau}|$
and $|\hat G_{e\tau}/\hat G_{e\mu}|$, respectively,
where
\be\label{eq:Ghat_def}
\widehat G_{l'l} \,\equiv\, G_{l'l} \,
\sqrt{1 + 4\,\eta_{\l\to l'}\, {m_{l'}\over m_l}\,
{g\!\left( m_{l'}^2/ m_l^2 \right)\over
f\!\left( m_{l'}^2/ m_l^2 \right)}}
\, .
\ee
An important point, emphatically stressed by
Fetscher and Gerber \cite{FG:93}, concerns the extraction
of $G_{e \mu}$, whose uncertainty is dominated
by the uncertainty in $\eta_{\mu\to e}$.

In terms of the $g_{\epsilon\omega}^n$
couplings, the shape parameters in Eq.~\eqn{eq:spectrum}
are:
\beqn\label{eq:michel}
\rho - \frac{3}{4} & \!\!\! = & \!\!\!
- \frac{3}{4}
\left[ {|g^V_{LR}|}^2 + {|g^V_{RL}|}^2 + 2 {|g^T_{LR}|}^2
+ 2{|g^T_{RL}|}^2 +
\mbox{\rm Re}(g^S_{LR} g^{T \ast}_{LR} + g^S_{RL} g^{T \ast}_{RL})
\right] ,
\nonumber\\
\eta
& \!\!\! = & \!\!\!
\frac{1}{2} \mbox{\rm Re}\left[
g^V_{LL} g^{S\ast}_{RR} + g^V_{RR}  g^{S\ast}_{LL}
+ g^V_{LR} \left(g^{S\ast}_{RL} + 6 g^{T\ast}_{RL}\right)
+ g^V_{RL} \left(g^{S\ast}_{LR} + 6 g^{T\ast}_{LR}\right)
\right] ,
\nonumber\\
\xi - 1
& \!\!\! = & \!\!\!
- \frac{1}{2}
\left[ {|g^S_{LR}|}^2 + {|g^S_{RR}|}^2
+ 4 (-{|g^V_{LR}|}^2 + 2 {|g^V_{RL}|}^2 + {|g^V_{RR}|}^2)
\right.
\nonumber\\
& &
\hspace{5mm}
\left.
- 4 {|g^T_{LR}|}^2 + 16 {|g^T_{RL}|}^2
- 8 {\rm Re}(g^S_{LR} g^{T \ast}_{LR} - g^S_{RL} g^{T \ast}_{RL})
\right]\ ,
\\
({\xi}\delta) - \frac{3}{4}
& \!\!\! = & \!\!\!
- \frac{3}{4}
\left[ \frac{1}{2} ({|g^S_{LR}|}^2 + {|g^S_{RR}|}^2)
+ ({|g^V_{LR}|}^2 + {|g^V_{RL}|}^2 + 2 {|g^V_{RR}|}^2)
\right.
\nonumber\\
&   &
\hspace{5mm}
\left.
+ 2 ({2 |g^T_{LR}|}^2 + {|g^T_{RL}|}^2)
- \mbox{\rm Re}(g^S_{LR} g^{T \ast}_{LR} - g^S_{RL} g^{T \ast}_{RL})
\right]\ . \nonumber
\eeqn
In the SM,  
$\rho = \delta = 3/4$,
$\eta = \eta'' = \alpha' = \beta' = 0 $ and
$\xi = \xi' = \xi'' = 1 $.

It is convenient to introduce \cite{FGJ:86} the probabilities
$Q_{\epsilon\omega}$ for the
decay of a $\omega$-handed $l^-$
into an $\epsilon$-handed
daughter lepton,
\begin{eqnarray}\label{eq:Q_LL}
Q_{LL} &\!\!\! = &\!\!\!
{1 \over 4} |g^S_{LL}|^2 \! +  |g^V_{LL}|^2
\phantom{+ 3 |g^T_{LR}|^2}
 = {1 \over 4}\left(
-3 +{16\over 3}\rho -{1\over 3}\xi +{16\over 9}\xi\delta +\xi'+\xi''
\right)\! , \quad\;\;\no\\
Q_{RR} &\!\!\! = &\!\!\!
{1 \over 4} |g^S_{RR}|^2 \! + \! |g^V_{RR}|^2
\phantom{+ 3 |g^T_{LR}|^2}
 =  {1 \over 4}\left(
-3 +{16\over 3}\rho +{1\over 3}\xi -{16\over 9}\xi\delta -\xi'+\xi''
\right)\!  , \no\\
Q_{LR} &\!\!\! = &\!\!\!
{1 \over 4} |g^S_{LR}|^2 \! + \!  |g^V_{LR}|^2
            \! + \!   3 |g^T_{LR}|^2
 = {1 \over 4}\left(
5 -{16\over 3}\rho +{1\over 3}\xi -{16\over 9}\xi\delta +\xi'-\xi''
\right)\! , \\
Q_{RL} &\!\!\! = &\!\!\!
{1 \over 4} |g^S_{RL}|^2  \! + \!  |g^V_{RL}|^2
            \! + \!  3 |g^T_{RL}|^2
= {1 \over 4}\left(
5 -{16\over 3}\rho -{1\over 3}\xi +{16\over 9}\xi\delta -\xi'-\xi''
\right)\! . \no
\end{eqnarray}
Upper bounds on any of these (positive-semidefinite) probabilities
translate into corresponding limits for all couplings with the
given chiralities.

For $\mu$-decay, where precise measurements of the polarizations of
both $\mu$ and $e$ have been performed, there exist \cite{FGJ:86}
upper bounds on $Q_{RR}$, $Q_{LR}$ and $Q_{RL}$, and a lower bound
on $Q_{LL}$. They imply corresponding upper bounds on the 8
couplings $|g^n_{RR}|$, $|g^n_{LR}|$ and $|g^n_{RL}|$.
The measurements of the $\mu^-$ and the $e^-$ do not allow us to
determine $|g^S_{LL}|$ and $|g^V_{LL}|$ separately \cite{FGJ:86,JA:66}.
Nevertheless, since the helicity of the $\nu_\mu$ in pion decay is
experimentally known \cite{RO:82}
to be $-1$, a lower limit on $|g^V_{LL}|$ is
obtained \cite{FGJ:86} from the inverse muon decay
$\nu_\mu e^-\to\mu^-\nu_e$.
The present (90\% CL) bounds \cite{PDG:94,BA:88}
on the $\mu$-decay couplings
are given in Table~\ref{tab:mu_couplings}. These limits show nicely
that the bulk of the $\mu$-decay transition amplitude is indeed of
the predicted V$-$A type.

\begin{table}[hbt]
\caption{90\% CL experimental limits \protect\cite{PDG:94,BA:88}
for the $\mu$-decay $g^n_{e_\epsilon \mu_\omega}$ couplings.}
\label{tab:mu_couplings}
\centering\vspace{0.2cm}
\begin{tabular}{|l|l|l|}
\hline
$|g^S_{e_L \mu_L}| < 0.55$  & $|g^V_{e_L \mu_L}| > 0.96$ &
\hfil -- \hfil \\
$|g^S_{e_R \mu_R}| < 0.066$ & $|g^V_{e_R \mu_R}| < 0.033$ &
\hfil -- \hfil \\
$|g^S_{e_L \mu_R}| < 0.125$ & $|g^V_{e_L \mu_R}| < 0.060$ &
 $|g^T_{e_L \mu_R}| < 0.036$\\
$|g^S_{e_R \mu_L}| < 0.424$ & $|g^V_{e_R \mu_L}| < 0.110$ &
 $|g^T_{e_R \mu_L}| < 0.122$\\
\hline
\end{tabular}
\end{table}

The experimental analysis of the $\tau$-decay parameters is
necessarily
different from the one applied to the muon, because of the much
shorter $\tau$ lifetime.
The measurement of the $\tau$ polarization and the parameters
$\xi$ and $\delta$ is still possible due to the fact that the spins
of the $\tau^+\tau^-$ pair produced in $e^+e^-$ annihilation
are strongly correlated
[27--34].
However,
the polarization of the charged lepton emitted in the $\tau$ decay
has never been measured. In principle, this could be done
for the decay $\tau^-\to\mu^-\bar\nu_\mu\nu_\tau$ by stopping the
muons and detecting their decay products \cite{FE:90}.
The measurement of the inverse decay $\nu_\tau l^-\to\tau^-\nu_l$
looks far out of reach.

The present experimental status on the $\tau$-decay Michel parameters
is shown in Table~\ref{tab:tau_michel} \cite{PS:95},
which gives the world-averages of all published
[3, 35--37]
measurements.
For comparison, the values measured in $\mu$-decay \cite{PDG:94}
are also given.
The improved accuracy of the most recent experimental analyses
has brought an enhanced sensitivity to the different shape parameters,
allowing the first measurements of $\eta_{\tau\to\mu}$
\cite{ALEPH:95b,ARGUS:95},
$\xi_{\tau\to e}$, $\xi_{\tau\to\mu}$, $(\xi\delta)_{\tau\to e}$ and
$(\xi\delta)_{\tau\to\mu}$ \cite{ALEPH:95b}.
(The ARGUS measurement \cite{ARGUS:95b} of $\xi_{\tau\to l}$ and
$(\xi\delta)_{\tau\to l}$ assumes identical couplings for $l=e,\mu$.
A measurement of $\sqrt{\xi_{\tau\to e}\xi_{\tau\to\mu}}$
was published previously \cite{ARGUS:93}).

\begin{table}[bt]   
\caption{Experimental averages
[3, 35--37]
of the Michel parameters.
The last column ($\tau\to l$) assumes identical couplings
for $l=e,\mu$
(the quoted value for $\eta_{\tau\to l}$ is that obtained directly
from measurements of the energy distribution).
$\xi_{\mu\to e}$ refers to the product $\xi_{\mu\to e}\cP_\mu$,
where $\cP_\mu\approx 1$ is the longitudinal polarization
of the muon from pion decay.}
\label{tab:tau_michel}
\centering\vspace{0.2cm}
\begin{tabular}{|c|c|c|c|c|}
\hline
& $\mu\to e$ & $\tau\to\mu$ & $\tau\to e$ & $\tau\to l$
\\ \hline
$\rho$ & $0.7518\pm 0.0026$ & $0.738\pm 0.038$ & $0.736\pm 0.028$ &
$0.733\pm 0.022$
\\
$\eta$ & $-0.007\pm 0.013$ & $-0.14\pm 0.23\phantom{-}$ & -- &
$-0.01\pm 0.14\phantom{-}$
\\
$\xi$ & $1.0027\pm 0.0085$ & $1.23\pm 0.24$ & $1.03\pm 0.25$ &
$1.06\pm 0.11$
\\
$\xi\delta$ & $0.7506\pm 0.0074$ & $0.71\pm 0.15$ & $ 1.11\pm 0.18$ &
$ 0.76\pm 0.09$
\\ \hline
\end{tabular}
\end{table}

The determination of the $\tau$-polarization
parameters \cite{ALEPH:95b,ARGUS:95b}   
allows us to bound the total probability for the decay of
a right-handed $\tau$ \cite{FE:90},
\be\label{eq:Q_R}
Q_{\tau_R} \equiv Q_{l'_R\tau^{\phantom{'}}_R} +
Q_{l'_L\tau^{\phantom{'}}_R}
= \frac{1}{2}\, \left[ 1 + \frac{\xi}{3} - \frac{16}{9}
(\xi\delta)\right]
\; .
\ee
One finds (ignoring possible correlations among the measurements)
\cite{PS:95}:
\begin{eqnarray}
Q_{\tau_R}^{\tau\to\mu} &\!\!\! =&\!\!\! \phantom{-}0.07\pm 0.14 \;
< \, 0.28 \quad (90\%\;\mbox{\rm CL})\, , \no\\
Q_{\tau_R}^{\tau\to e} &\!\!\! =&\!\!\! -0.32\pm 0.17 \;
< \, 0.14 \quad (90\%\;\mbox{\rm CL})\, , \\
Q_{\tau_R}^{\tau\to l} &\!\!\! =&\!\!\! \phantom{-}0.00\pm 0.08 \;
< \, 0.14 \quad (90\%\;\mbox{\rm CL})\, , \no
\end{eqnarray}
where the last value refers to the $\tau$-decay into either $l=e$ or $\mu$,
assuming universal leptonic couplings.
Since these probabilities are positive semidefinite quantities, they imply
corresponding limits on all $|g^n_{l_R\tau_R}|$
and $|g^n_{l_L\tau_R}|$ couplings.
The quoted 90\% CL have been obtained
adopting a Bayesian approach for one-sided limits \cite{PDG:94}.
Table~\ref{table:g_tau_bounds} gives the implied bounds on the
$\tau$-decay couplings.

\begin{table}[bt]
\caption{90\% CL limits \protect\cite{PS:95}
for the $\tau_R$-decay $g^n_{l_\epsilon \tau_R}$ couplings.
The numbers with an asterisk use the measured value of
$(\xi\delta)_e$.}
\label{table:g_tau_bounds}
\centering\vspace{0.2cm}
\begin{tabular}{|l|l|l|}
\hline
\hfil $\tau\to\mu$\hfil &\hfil $\tau\to e$ \hfil &
\hfil $\tau\to l$ \hfil
\\\hline
$|g^S_{\mu_R\tau_R}| < 1.05$ &
$|g^S_{e_R\tau_R}| < 0.75^*$ &
$|g^S_{l_R\tau_R}| < 0.74$
\\
$|g^S_{\mu_L\tau_R}| < 1.05$ &
$|g^S_{e_L\tau_R}| < 0.75^*$ &
$|g^S_{l_L\tau_R}| < 0.74$
\\ \hline
$|g^V_{\mu_R\tau_R}| < 0.53$ &
$|g^V_{e_R\tau_R}| < 0.38^*$ &
$|g^V_{l_R\tau_R}| < 0.37$
\\
$|g^V_{\mu_L\tau_R}| < 0.53$ &
$|g^V_{e_L\tau_R}| < 0.38^*$ &
$|g^V_{l_L\tau_R}| < 0.37$
\\ \hline
$|g^T_{\mu_L\tau_R}| < 0.30$ &
$|g^T_{e_L\tau_R}| < 0.22^*$ &
$|g^T_{l_L\tau_R}| < 0.21$
\\ \hline
\end{tabular}
\end{table}

The central value of $Q_{\tau_R}^{\tau\to e}$
turns out
to be negative at the $2\sigma$ level; i.e., there is only a 3\%
probability to have a positive value of
$Q_{\tau_R}^{\tau\to e}$. Therefore, the limits on
$|g^n_{e_R\tau_R}|$ and $|g^n_{e_L\tau_R}|$
should be taken with some caution, since the meaning of the assigned
confidence level is not at all clear.
The problem clearly comes from the measured value of $(\xi\delta)_e$.
In order to get a positive probability
$Q_{\tau_R}$, one needs
$(\xi -1) > \frac{16}{3} [(\xi\delta) -\frac{3}{4}]$.
Thus, $(\xi\delta)$ can only be made larger than
$3/4$ at the expense of making $\xi$ correspondingly much
larger than one \cite{PS:95}.

If lepton universality is assumed (i.e. $G_{l'l} = G_F$,
$\, g^n_{l'_\epsilon l^{\phantom{'}}_\omega}\! = g^n_{\epsilon\omega}$),
the leptonic decay ratios $B_\mu/B_e$  and $B_e\tau_\mu/\tau_\tau$
provide limits on the low-energy parameter $\eta$.
The best sensitivity \cite{ST:94} comes from
$\widehat G_{\mu\tau}$,
where the term proportional to $\eta$ is not suppressed by
the small $m_e/m_l$ factor. The measured $B_\mu/B_e$ ratio implies
then \cite{PS:95}:
\be\label{eq:eta_univ}
\eta_{\tau\to l} \, = \, 0.007\pm 0.033  \ .
\ee
This determination is more accurate that the one in
Table~\ref{tab:tau_michel},
obtained from the shape of the energy distribution,
and is comparable to the value measured in $\mu$-decay:
$\eta_{\mu\to e} = -0.007\pm 0.013$ \cite{BU:85}.

A non-zero value of $\eta$ would show that there are at least two
different couplings with opposite chiralities for the charged leptons.
Since, we assume the V$-$A coupling $g_{LL}^V$ to be dominant, the
second coupling would be \cite{FE:90} a Higgs-type coupling $g^S_{RR}$
[$\eta\approx\mbox{\rm Re}(g^S_{RR})/2$,
to first-order in new-physics contributions].
Thus, Eq.~(\ref{eq:eta_univ}) puts the (90\% CL) bound:
$-0.09 \, <\mbox{\rm Re}(g^S_{RR}) < 0.12$.

\subsection*{\centering Model-Dependent Constraints}

The general bounds in Table~\ref{table:g_tau_bounds} look
rather weak. The sensitivity of present experiments is not
good enough to get interesting constraints from a completely
general analysis of the four-fermion Hamiltonian. 
Nevertheless, stronger limits can be obtained within particular
models, as shown in Tables~\ref{tab:coup_CH}, \ref{tab:W_couplings}
and \ref{tab:ns_coup}.

Table~\ref{tab:coup_CH} assumes that there are no tensor couplings,
i.e. $g^T_{\epsilon\omega}=0$. This condition is satisfied in
any model where the interactions are mediated by vector bosons
and/or charged scalars \cite{PS:95}.
In this case, 
the quantities  $(1-\frac{4}{3}\rho)$,
$(1-\frac{4}{3}\xi\delta)$ and
$(1-\frac{4}{3}\rho) + \frac{1}{2} (1-\xi)$
reduce to sums of $|g^n_{l'_\epsilon l_\omega}|^2$,
which are positive semidefinite;
i.e.~,
in the absence of tensor couplings, $\rho\leq\frac{3}{4}$,
$\xi\delta\leq\frac{3}{4}$
and $(1-\xi) > 2 (\frac{4}{3}\rho - 1)$.
This allows us to extract direct bounds on
several couplings.

\begin{table}[bth]
\caption{90\% CL limits
for the couplings $g^n_{\epsilon\omega}$, assuming that there are no
tensor couplings \protect\cite{PS:95}.
The numbers with an asterisk use the measured value of $(\xi\delta)_e$.}
\label{tab:coup_CH}
\centering\vspace{0.2cm}
\begin{tabular}{|l|l|l|l|l|}
\hline & \hfil $\mu\to e$\hfil &
\hfil $\tau\to\mu$\hfil &\hfil $\tau\to e$ \hfil &
\hfil $\tau\to l$ \hfil
\\\hline
$|g^S_{LL}|$ & $<0.55$ & $\leq 2$ & $\leq 2$ & $\leq 2$
\\
$|g^S_{RR}|$ & $<0.066$ & $<0.80$ & $<0.63^*$ & $<0.62$
\\
$|g^S_{LR}|$ & $<0.125$ & $<0.80$ & $<0.63^*$ & $<0.62$
\\
$|g^S_{RL}|$ & $<0.424$ & $\leq 2$ & $\leq 2$ & $\leq 2$
\\ \hline
$|g^V_{LL}|$ & $>0.96$ & $\leq 1$ & $\leq 1$ & $\leq 1$
\\
$|g^V_{RR}|$ & $<0.033$ & $<0.40$ & $<0.32^*$ & $<0.31$
\\
$|g^V_{LR}|$ & $<0.060$ & $<0.31$ & $<0.27$ & $<0.25$
\\
$|g^V_{RL}|$ & $<0.047$ & $<0.23$ & $<0.27$ & $<0.18$
\\ \hline
\end{tabular}
\end{table}

If one only considers $W$-mediated interactions, but admitting the
possibility that the $W$ couples non-universally to leptons of any
chirality, the stronger limits in
Table~\ref{tab:W_couplings} are obtained \cite{PS:95}.
In this case, the $g^V_{l'_\epsilon l^{\phantom{'}}_\omega}$
constants factorize into the product of two leptonic $W$ couplings,
implying \cite{MU:85} additional relations among the couplings,
such as $g^V_{LR}\ g^V_{RL} = g^V_{LL}\ g^V_{RR}$, which hold within
any of the three channels, $(\mu, e)$,
$(\tau, e)$, and $(\tau, \mu)$.
Moreover,
there are additional equations relating different processes,
such as \cite{PS:95}
$g^V_{\mu_L \tau_L}\ g^V_{e_L \tau_R}  =
g^V_{\mu_L \tau_R}\ g^V_{e_L \tau_L}$.
The normalization condition~\eqn{eq:normalization}
provides lower bounds on the $g^V_{LL}$ couplings.

\begin{table}[bt]
\caption{90\% CL limits  on the $g^V_{\epsilon \omega}$ couplings,
assuming that  (non-standard) $W$-exchange is the only relevant interaction
\protect\cite{PS:95}.}
\label{tab:W_couplings}
\centering\vspace{0.2cm}
\begin{tabular}{|l|l|l|l|}
\hline & \hfil $\mu\to e$\hfil &
\hfil $\tau\to\mu$\hfil &\hfil $\tau\to e$ \hfil
\\\hline
$|g^V_{LL}|$ & $>0.997$ & $>0.95$ & $>0.96$
\\
$|g^V_{RR}|$ & $<0.0028$ & $<0.019$ & $<0.013$
\\
$|g^V_{LR}|$ & $<0.060$ & $<0.31$ & $<0.27$
\\
$|g^V_{RL}|$ & $<0.047$ & $<0.060$ & $<0.047$
\\ \hline
\end{tabular}
\end{table}

For $W$-mediated interactions,
the hadronic $\tau$-decay modes can also be used to test the
structure of the $\tau\nu_\tau W$
vertex, if one assumes that the W coupling to the light quarks
is the SM one.
The $\cP_\tau$ dependent part of the decay amplitude is then
proportional to the mean $\nu_\tau$ helicity
\bel{eq:nu_helicity}
h_{\nu_\tau} \,\equiv\, {|g_R|^2 - |g_L|^2
\over |g_R|^2 + |g_L|^2} ,
\ee
which plays a role analogous to the leptonic-decay parameter $\xi$.
The analysis of $\tau^+\tau^-$ decay correlations
in leptonic--hadronic and hadronic--hadronic decay modes, using
the $\pi$, $\rho$ and $a_1$ hadronic final states, gives
$h_{\nu_\tau}=-1.014\pm 0.027$
[35,37,42];
this implies $|g_R/g_L|^2= -0.007\pm 0.013 < 0.018$ (90\% CL).
The sign of the $\nu_\tau$ helicity can be determined \cite{ARGUS:90}
to be negative
with the decay $\tau^-\nu_\tau a_1^-$, because there are
two different amplitudes [corresponding to two different ways of
forming the rho in $a_1^-\to(\rho\pi)^-$]
and their interference contains
information on the sign.\footnote{
Once the $h_{\nu_\tau}$ sign is fixed, the measurement
of leptonic--hadronic correlations determines the signs of
$\xi_{\tau\to e}$ and $\xi_{\tau\to\mu}$ to be positive.
At the $Z$ peak, the signs of $\xi_{\tau\to l}$ and $h_{\nu_\tau}$
can be directly determined \protect\cite{ALEPH:94}
from the sign of $\cP_\tau$, which is
fixed by combining the measurements of the
polarization and left-right asymmetries.}

\begin{table}[tbh]
\caption{90\% CL limits
for the $g^n_{\epsilon\omega}$ couplings, taking
$g^n_{RR}=0$, $g^S_{LL}=0$, $g^V_{LR}=g^S_{LR}=2 g^T_{LR}$ and
$g^V_{RL}=g^S_{RL}=2 g^T_{RL}$ \protect\cite{PS:95}.}
\label{tab:ns_coup}
\centering\vspace{0.2cm}
\begin{tabular}{|l|l|l|l|l|}
\hline & \hfil $\mu\to e$\hfil &
\hfil $\tau\to\mu$\hfil &\hfil $\tau\to e$ \hfil &
\hfil $\tau\to l$ \hfil
\\\hline
$|g^V_{LL}|$ & $>0.998$ & $>0.95$ & $>0.96$ & $>0.97$
\\
$|g^V_{LR}|$ & $<0.047$ & $<0.22$ & $<0.19$ & $<0.18$
\\
$|g^V_{RL}|$ & $<0.033$ & $<0.16$ & $<0.19$ & $<0.13$
\\ \hline
\end{tabular}
\end{table}

Table~\ref{tab:ns_coup} shows the constraints obtained under
the assumption that the interaction is mediated by the SM $W$
plus an additional neutral scalar \cite{PS:95}.
The scalar contributions vanish for the LL and RR couplings
and satisfy the relations
$g^V_{LR}  = g^S_{LR} = 2 g^T_{LR}$,
$g^V_{RL}  = g^S_{RL} = 2 g^T_{RL}$. This allows to
express everything in terms of the vector couplings.
The quantities $(1-\frac{4}{3}\rho)$,
$(1-\frac{4}{3}\xi\delta)$ and
$(1-\frac{4}{3}\rho) + \frac{1}{2} (1-\xi)$
are also positive semidefinite in this case. Moreover,
$(1-\frac{4}{3}\rho)=(1-\frac{4}{3}\xi\delta)$.

\subsection*{\centering Expected Signals in Minimal New-Physics Scenarios}

All experimental results obtained so far are consistent with the
SM. Clearly, the SM
provides the dominant contributions to the $\tau$-decay amplitudes.
Future high-precision measurements of allowed $\tau$-decay modes
should then look for small deviations of the SM predictions and
find out the possible source of any detected discrepancy.

In a first analysis, it seems natural to assume \cite{PS:95}
that new-physics
effects would be dominated by the exchange of a single
intermediate boson, coupling to two leptonic currents.
The new contribution could be originated by non-standard couplings
of the usual $W$ boson, or by the exchange of
a new scalar or vector particle (intermediate tensor particles hardly
appear in any reasonable model beyond the SM).

Table~\ref{tab:summary} \cite{PS:95}
summarizes the expected effects of different new-physics scenarios
on the measurable shape parameters.
The four general cases studied correspond to adding a single intermediate
boson-exchange, $V^+$, $S^+$, $V^0$, $S^0$
(charged/neutral, vector/scalar), to the SM contribution
(a non-standard $W$ would be a particular case of the
SM + $V^+$ scenario).
AS indicates that any sign is allowed.

\begin{table}[bht]
\caption{Theoretical constraints
on the Michel parameters \protect\cite{PS:95}}
\label{tab:summary}
\centering\vspace{0.2cm}
\begin{tabular}{|c|c|c|c|c|}
\hline
&  SM + $V^+$  &  SM + $S^+$  &  SM + $V^0$  &  SM + $S^0$
\\ \hline
$\rho - 3/4$     & $< 0$  &  0   &  0   &  $< 0$
\\ \hline
$\xi - 1$        &   AS   & $< 0$ & $< 0$ &  AS
\\ \hline
$(\delta\xi)-3/4$&  $< 0$ & $< 0$ & $< 0$ & $< 0$
\\ \hline
$\eta$           &   0   &  AS   &   AS  &  AS
\\ \hline
\end{tabular}
\end{table}

It is immediately apparent that $\rho \leq 3/4$ and
$(\delta \xi) < 3/4$ in all cases studied. Thus one can have
new physics and still $\rho$ be equal to the SM value. In fact, any
interaction consisting of an arbitrary combination of $g^S_{\epsilon
\omega}$'s and $g^V_{\gamma \gamma}$'s yields this result
\cite{FE:90}. On the other hand, $(\delta\xi)$ will be different from
$3/4$ in any of the cases above providing, in principle, a better
opportunity for the detection of Physics Beyond the SM.

The above features are easy to understand by looking back at
Eqs.~(\ref{eq:michel}) and recalling that the tensor couplings can only be
generated by neutral scalar interactions (violating individual lepton
flavours), in which case they are proportional to the scalar couplings. It
is easy to see that having two such neutral scalars will not alter the
situation. Indeed, to obtain
$\rho > 3/4$ or $(\delta \xi) > 3/4$ one would need
to get contributions from
charged and neutral scalars simultaneously\cite{PS:95}.
Moreover, $(\delta \xi)>3/4$ can only happen through $RL$ couplings
and must be accompanied by $\xi>1$.

The \tcf\ offers an ideal experimental environment to perform
this kind of analyses. The  expected sensitivities to the different
shape parameters, quoted in Table~\ref{tab:improvements}, would
allow to prove the effective four-fermion Hamiltonian to a level
where very interesting constraints on new-physics scenarios
could be obtained. The numbers given in Table~\ref{tab:improvements}
are somehow conservative, since they only take into account the
information obtained from correlated $\tau^+\tau^-$ events
where both $\tau$'s decay into leptons \cite{marbella:1,ST:93}.
Better precisions may be
reached including  the correlations of the leptonic decays with
the hadronic ones \cite{marbella:1,ST:93}.

\section*{\centering DISCUSSION}

The flavour structure of the SM is one of the main pending questions
in our understanding of weak interactions.
Although we do not know the reason of the observed
family replication, we have learn experimentally that the
number of SM fermion generations is just three (and no more).
Therefore, we must study as precisely as possible
the few existing flavours,
to get some hints on the dynamics responsible for their
observed structure.
The construction of high-precision
flavour factories is clearly needed.

Without any doubt, the \tcf\ is the best available tool to explore
the $\tau$ and $\nu_\tau$ leptons and the charm quark.
This facility combines the three ingredients required
for making an accurate and exhaustive investigation of these
particles:
high statistics, low backgrounds and good control of systematic errors.
The threshold region provides a series of unique features
(low and measurable backgrounds free from heavy flavour contaminations,
monochromatic particles from two-body decays, small radiative
corrections, single tagging, high-rate calibration sources, \ldots)
that create an ideal experimental environment for this physics.

Two basic properties make the $\tau$ particle an ideal laboratory for
testing the SM: the $\tau$ is a lepton, which means clean physics,
and moreover, it is heavy enough to produce a large variety of decay
modes.
In the previous sections I have discussed two particular topics,
charged-current universality and Lorentz structure of the weak
currents, which would greatly benefit from a high-precision
experimental study of the $\tau$ lepton.
There are, in addition, many other interesting subjects to be
investigated.

The \tcf\ could carry out a precise and exhaustive study of all exclusive
$\tau$ decay channels, looking for signs of discrepancies with
the theoretical expectations.
The accurate measurement of the $q^2$ distribution of the final
hadrons would allow a detailed analysis of the vector and
axial-vector spectral functions and, therefore, a significant
improvement of our knowledge of QCD.
Rare and forbidden $\tau$ decays could be looked for, with a sensitivity
better than $10^{-7}$ in some channels.
The bound on the $\nu_\tau$ mass could be pushed down to the
1--2 MeV level.
The present knowledge of the $\tau$ electromagnetic moments
could be improved by more than one order of magnitude.
Last but not least, CP-violation in the lepton sector at the
milli-weak ($10^{-3}$) level could be investigated
(with longitudinal beam polarization).

In addition to the large improvement in our knowledge of the
$\tau$ lepton, the \tcf\ would also provide precious information
on the $c$ quark, through the detailed study of the $D$ mesons and the
$J/\Psi$ and other charmonium states.
A comprehensive set of precision measurements for $\tau$, charm and
light-hadron spectroscopy would be obtained, proving the
SM to a much deeper level of sensitivity and exploring the frontiers
of its possible extensions.

\section*{\centering ACKNOWLEDGEMENTS}

Many results discussed here have been obtained in collaboration
with Jo\~ao P.~Silva \cite{PS:95}.
This work has been supported in part by CICYT (Spain),
under grant No. AEN-93-0234.

\begin{Thebibliography}{99}

\refjl{PDG:90}{M. Aguilar-Ben\'{\i}tez \etal}{Review of Particle
   Properties, \PL}{B239}{1990}{1}

\bibitem{PI:92} A. Pich, {\it Tau Physics}, in {\it Heavy Flavours},
  eds. A.J.~Buras and M.~Lindner,
  Advanced Series on Directions in High Energy Physics -- Vol.~10
  (World Scientific, Singapore, 1992), p.~375.

\refjl{PDG:94}{M. Aguilar-Ben\'{\i}tez \etal}{Review of Particle
   Properties, \PR}{D50}{1994}{1173}

\bibitem{montreux} Proc. {\it Third Workshop on Tau Lepton Physics}
 (Montreux, 1994), ed. L.~Rolandi, {\it \NP\ B (Proc. Suppl.)\/} {\bf 40}
 (1995).

\bibitem{marbella:1} A. Pich, {\it Perspectives on $\tau$-Charm Factory
  Physics}, in Proc. {\it Third Workshop on the Tau-Charm Factory}
  (Marbella, 1993), eds. J.~Kirkby and R.~Kirkby (Editions Fronti\'eres,
  Gif-sur-Yvette, 1994), p.~767.

\bibitem{marbella:2} A. Pich, {\it Tau Physics Prospects at the
  $\tau$-Charm Factory and at other Machines},
  in Proc. {\it Third Workshop on the Tau-Charm Factory}
  (Marbella, 1993), eds. J.~Kirkby and R.~Kirkby (Editions Fronti\'eres,
  Gif-sur-Yvette, 1994), p.~51.

\bibitem{QCD:94} A. Pich, {\it QCD Predictions for the $\tau$ Hadronic
   Width: Determination of $\alpha_s(M_\tau^2)$}, in Proc.
   {\it QCD 94} (Montpellier, 1994), ed. S. Narison,
   {\it \NP\ B (Proc. Suppl.) \/} {\bf 39B,C} (1995) 326.

\bibitem{slac:89} A. Pich, {\it QCD Tests from Tau Decay Data}, in
    Proc. {\it Tau-Charm Factory Workshop} (SLAC, California,  1989),
    ed. L.V.~Beers, SLAC-Report-343 (1989), p.~416.

\refjl{MS:88}{W.J. Marciano and A. Sirlin}{\PRL}{61}{1988}{1815}

\refjl{PDG:92}{M. Aguilar-Ben\'{\i}tez \etal}{Review of Particle
    Properties, \PR}{D45}{1992}{Part 2}

\refjl{ALEPH:95}{D. Buskulic \etal\ (ALEPH)}{\PL}{B349}{1995}{585}

\refjl{BR:92}{D.I. Britton \etal}{\PRL}{68}{1992}{3000; \\
  G. Czapek \etal {\it\PRL} {\bf 70} (1993) 17.}

\bibitem{UA1:89}
   C. Albajar \etal\ (UA1), {\it \ZP} {\bf C44} (1989) 15; \\
   J. Alitti \etal\ (UA2), {\it \PL} {\bf B280} (1992) 137; \\
   F. Abe \etal\ (CDF), {\it \PRL} {\bf 68} (1992) 3398;
      {\bf 69} (1992) 28.

\refjl{MS:93}{W.J. Marciano and A. Sirlin}{\PRL}{71}{1993}{3629}

\bibitem{DF:94}
   R. Decker and M. Finkemeier, {\it\NP} {\bf B438} (1995) 17;
   {\it\NP\ B (Proc. Suppl.)} {\bf 40} (1995) 453.

\refjl{MA:94}{W.J. Marciano}{\NP\ B (Proc. Suppl.)}{40}{1995}{3}

\refjl{MI:50}{L. Michel}{Proc. Phys. Soc.}{A63}{1950}{514; 1371}

\refjl{BM:57}{C. Bouchiat and L. Michel}{\PR}{106}{1957}{170}

\refjl{KS:57}{T. Kinoshita and A. Sirlin}{\PR}{107}{1957}{593;
     {\bf 108} (1957) 844}

\bibitem{SCH:83} F. Scheck, {\it Leptons, Hadrons and Nuclei}
   (North-Holland, Amsterdam, 1983);
   {\it\PRep} {\bf 44} (1978) 187.

\refjl{FGJ:86}{W. Fetscher, H.-J. Gerber and K.F. Johnson}{\PL}{B173}
    {1986}{102}

\bibitem{FG:93} W. Fetscher and H.-J. Gerber, {\it Precision
  Measurements in Muon and Tau Decays}, in {\it Precision Tests of the
  Standard Electroweak Model}, ed. P.~Langacker,
  Advanced Series on Directions in High Energy Physics -- Vol.~14
  (World Scientific, Singapore, 1995), p.~657.

\refjl{PS:95}{A. Pich and J.P. Silva}{\PR}{D}{1995}{in press
  [hep-ph/9505327]}

\refjl{JA:66}{C. Jarlskog}{\NP}{75}{1966}{659}

\bibitem{RO:82}
   L.Ph. Roesch \etal , {\it Helv. Phys. Acta} {\bf 55} (1982) 74; \\
   W. Fetscher, {\it\PL} {\bf 140B} (1984) 117; \\
   A. Jodidio \etal , {\it\PR} {\bf D34} (1986) 1967;
     {\it\PR} {\bf D37} (1988) 237.

\bibitem{BA:88}
   B. Balke \etal , {\it\PR} {\bf D37} (1988) 587; \\
   D. Geiregat \etal , {\it\PL} {\bf B247} (1990) 131; \\
   S.R. Mishra \etal , {\it\PL} {\bf B252} (1990) 170.

\refjl{TS:71}{Y.S. Tsai}{\PR}{D4}{1971}{2821; \\
%
   S. Kawasaki, T. Shirafuji and S.Y. Tsai, {\it Progr. Theor. Phys.}
   {\bf 49} (1973) 1656}

\refjl{PS:77}{S.-Y. Pi and A.I. Sanda}{\APNY}{106}{1977}{171}

\bibitem{GO:89} J.J. G\'omez-Cadenas, {\it Beautiful $\tau$ Physics
in the Charm Land}, in Proc. {\it Tau-Charm Factory
 Workshop} (SLAC, 1989), ed. L.V. Beers, SLAC-Report-343 (1989), p.~48.

\bibitem{NE:91} C.A. Nelson, {\it\PR} {\bf D43} (1991) 1465;
   {\it\PRL} {\bf 62} (1989) 1347; {\it\PR} {\bf D40} (1989) 123
    [Err: {\bf D41} (1990) 2327]; \\
%
   S. Goozovat and C.A. Nelson, {\it\PR} {\bf D44} (1991) 2818;
   {\it\PL} {\bf B267} (1991) 128 [Err: {\bf B271} (1991) 468].

\refjl{FE:90}{W. Fetscher}{\PR}{D42}{1990}{1544}

\refjl{BPR:91}{J. Bernab\'eu, A. Pich and N. Rius}{\PL}{B257}{1991}{219}

\refjl{ABGPR:92}{R. Alemany \etal}{\NP}{B379}{1992}{3}

\refjl{DDDR:93}{M. Davier \etal}{\PL}{B306}{1993}{411}

\refjl{ALEPH:95b}{D. Buskulic \etal\ (ALEPH)}{\PL}{B346}{1995}{379}

\refjl{ARGUS:95}{H. Albrecht \etal\ (ARGUS)}{\PL}{B341}{1995}{441}

\refjl{ARGUS:95b}{H. Albrecht \etal\ (ARGUS)}{\PL}{B349}{1995}{576}

\refjl{ARGUS:93}{H. Albrecht \etal\ (ARGUS)}{\PL}{B316}{1993}{608}

\refjl{ST:94}{A. Stahl}{\PL}{B324}{1994}{121}

\refjl{BU:85}{H. Burkard \etal}{\PL}{160B}{1985}{343}

\bibitem{MU:85}
  K. Mursula and F. Scheck, {\it\NP} {\bf B253} (1985) 189; \\
  K. Mursula, M. Roos and F. Scheck, {\it\NP} {\bf B219} (1983) 321.

\refjl{ARGUS:94}{H. Albrecht \etal\ (ARGUS)}{\PL}{B337}{1994}{383;
  \\ {\it\ZP} {\bf C58} (1993) 61}

\refjl{ARGUS:90}{H. Albrecht \etal\ (ARGUS)}{\PL}{B250}{1990}{164}

\refjl{ALEPH:94}{D. Buskulic \etal\ (ALEPH)}{\PL}{B321}{1994}{168}

\bibitem{ST:93} A. Stahl, {\it The Lorentz Structure of the Charged
  Weak Current in $\tau$ Decays},
  in Proc. {\it Third Workshop on the Tau-Charm Factory}
  (Marbella, 1993), eds. J. Kirkby and R. Kirkby (Editions Fronti\'eres,
  Gif-sur-Yvette, 1994), p.~175.

\end{Thebibliography}

\end{document}